\documentclass{elsart}
\usepackage{amssymb}
\usepackage{amsmath}
\usepackage{graphicx}

\begin{document}

\title{Symmetry breaking in a two-component system with repulsive
interactions and linear coupling}
\author{Hidetsugu Sakaguchi$^{1}$ and Boris A.Malomed$^{2,3}$}

\begin{abstract}
We extend the well-known theoretical treatment of the spontaneous symmetry
breaking (SSB) in two-component systems combining linear coupling and
self-attractive nonlinearity to a system in which the linear coupling
competes with repulsive interactions. First, we address one- and
two-dimensional (1D and 2D) ground-state (GS) solutions and 2D vortex states
with topological charges $S=1$ and $2$, maintained by a confining
harmonic-oscillator (HO) potential. The system can be implemented in BEC and
optics. By means of the Thomas-Fermi approximation and numerical solution of
the underlying coupled Gross-Pitaevskii equations, we demonstrate that SSB
takes place, in the GSs and vortices alike, when the cross-component
repulsion is stronger that the self-repulsion in each component. The SSB
transition is categorized as a supercritical bifurcation, which gives rise
to states featuring broken symmetry in an inner area, and intact symmetry in
a surrounding layer. Unlike stable GSs and vortices with $S=1$, the states
with $S=2$ are unstable against splitting. We also address SSB for 1D gap
solitons in the system including a lattice potential. In this case, SSB
takes place under the opposite condition, i.e., the cross-component
repulsion must be weaker than the self-repulsion, and SSB is exhibited by
antisymmetric solitons.
\end{abstract}

\maketitle

\address{$^{1}$Department of Applied Science for Electronics and Materials,\\
Interdisciplinary Graduate School of Engineering Sciences,\\
Kyushu University, Kasuga, Fukuoka 816-8580, Japan\\
$^{2}$Department of Physical Electronics, School of Electrical Engineering,\\
Faculty of Engineering, and Center for Light-Matter Interaction,\\
Tel Aviv University, Tel Aviv 69978, Israel\\
$^{3}$Instituto de Alta Investigaci\'{o}n, Universidad de Tarapac\'{a},
Casilla 7D, Arica, Chile}

\section{Introduction and the model}

Spontaneous symmetry breaking (SSB) is a broad class of effects occurring in
systems combining spatial or inter-component symmetry and intrinsic
nonlinearity \cite{book}. While in linear systems the ground state (GS)
exactly reproduces the underlying symmetry \cite{LL}, and the GS is always a
single (non-degenerate) state, self-attraction in binary systems drives a
phase transition which destabilizes the symmetric GS, replacing it by a pair
of asymmetric ones that are mirror images of each other (so as to maintain
the overall symmetry). The transition takes place when the nonlinearity
strength attains a certain critical value. Starting from early works \cite%
{early1,early2}, many realizations of the SSB phenomenology have been
supplied by nonlinear optics and studies of Bose-Einstein condensates
(BECs). The similarity of these areas is underlain by the fact that the
nonlinear Schr\"{o}dinger equations \cite{NLS} and Gross-Pitaevskii
equations (GPEs) \cite{BEC1,BEC2}), which are basic models for optics and
BEC, respectively, are essentially identical, with the difference that the
evolution variable is the propagation distance, $z$, in optical waveguides,
and time, $t$, in BEC. In both cases SSB effects have been predicted in
two-components systems, with linear coupling between the components and
self-attraction in each component. In optics, a natural realization of the
\textit{nonlinear couplers} and SSB phenomenology in them is offered, in
terms of the temporal-domain propagation, by dual-core optical fibers. This
setting was studied in detail theoretically \cite{Trillo}-\cite{opt-coupler9}
(see also review \cite{Peng-book}), and nonlinear switching in the couplers
was demonstrated experimentally \cite{Denz,Bugar}. The same model may be
realized in the spatial domain, considering a planar dual-core nonlinear
waveguide \cite{Jensen,opt-coupler5,Raymond}. Another class of
nonlinear-optical symmetric systems and SSB dynamics in them, which were
explored both theoretically and experimentally, is provided by dual-cavity
lasers \cite{lasers1}-\cite{lasers4}.

A similar BEC\ system may be realized by loading the condensate into a pair
of parallel tunnel-coupled elongated traps, filled by self-attractive BEC
and linearly coupled by tunneling of atoms \cite{Marek}-\cite{Luca1}. SSB
effects, chiefly for matter-wave solitons, have been predicted in such dual
traps.

A more general two-component system includes, in addition to the
self-interaction of each component and linear mixing between them, nonlinear
cross-component interaction \cite{we1,Abad}. In optics, this model applies
to the propagation of electromagnetic waves with orthogonal circular
polarizations in the fiber with linear-polarization birefringence (that may
be induced by ellipticity of the fiber's cross section), which induces the
linear mixing \cite{birefr}. In BEC, the realization is possible in a single
elongated trap filled by a mixture of two different atomic states in the
condensate, while the linear (Rabi) mixing is imposed by a radiofrequency
field which resonantly couples the states \cite{Rabi1}-\cite{Merhasin}.

A majority of the above-mentioned works addressed effectively
one-dimensional (1D) settings. In optics, 2D two-component spatiotemporal
propagation may be realized in planar dual-core waveguides \cite{NDror}, and
similar realizations were considered for BEC loaded in dual-core
``pancake-shaped" traps \cite{Arik,sala-boris,Viskol}. The more general
system, which includes the inter-component nonlinearity, can be naturally
implemented as the spatial-domain propagation in bulk optical waveguides,
with the linear mixing of orthogonal circular polarizations induced by
intrinsic anisotropy (linear birefringence) in the transverse plane. In BEC,
an implementation is offered by a pancake trap filled by a binary
condensate, with the radiofrequency-induced coupling between two atomic
states.

As said above, the SSB of GSs in previous works was driven by attractive
nonlinearity. Self-repulsion does not break the GS symmetry, but it may
drive spontaneous breaking of antisymmetry of the first excited state, an
example being an antisymmetric bound state of matter-wave gap solitons in
tunnel-coupled traps equipped with periodic potentials \cite%
{Arik-1D,Arik,Marek2}. The objective of the present work is to demonstrate
that, nevertheless, the symmetry of the GS in 1D and 2D linearly-coupled
systems with repulsive nonlinearity may be spontaneously broken, replacing
it by a pair of asymmetric GSs, provided that the inter-component repulsion
dominates over the intra-component nonlinearity, and the GS is made
localized by a confining potential. This mechanism is similar to other
effects in which the dominant repulsion between two interacting fields (in
the absence of the linear coupling between them) makes their uniformly mixed
state unstable, well-known examples being phase separation in binary
mixtures \cite{Mineev} and the modulational instability in bimodal systems
in which the cross-repulsion is stronger than self-repulsion \cite{Agrawal}.

In terms of BEC, the system is represented by coupled GPEs for wave
functions $\phi _{1,2}$ of the two components, written in the 2D form, which
includes the confining harmonic-oscillator (HO) potential with strength $%
\Omega ^{2}$, acting in the $\left( x,y\right) $ plane, and scaled atomic
mass $m$:%
\begin{gather}
i\frac{\partial \phi _{1}}{\partial t}=-\frac{1}{2m}\nabla ^{2}\phi
_{1}+(g|\phi _{1}|^{2}+\gamma |\phi _{2}|^{2})\phi _{1}+\frac{\Omega ^{2}}{2}%
\left( x^{2}+y^{2}\right) \phi _{1}-\epsilon \phi _{2}  \label{CGL1} \\
i\frac{\partial \phi _{2}}{\partial t}=-\frac{1}{2m}\nabla ^{2}\phi
_{1}+(g|\phi _{2}|^{2}+\gamma |\phi _{1}|^{2})\phi _{2}+\frac{\Omega ^{2}}{2}%
\left( x^{2}+y^{2}\right) \phi _{2}-\epsilon \phi _{1}.  \label{CGL2}
\end{gather}%
Here, $\epsilon $ is the coefficient of the linear coupling, while $\gamma >0
$ and $g>0$ account for, severally, the cross-repulsion and self-repulsion.
These coefficients are subject to the above-mentioned condition of the
dominant cross-interaction, $\gamma >g$, except for Section IV (the case of $%
g<0$, i.e., self-attractive nonlinearity, may be considered too, although it
is less interesting than the case of the competition between the cross- and
self-repulsion). By means of obvious rescaling of $t$, coordinates, and wave
functions we fix
\begin{equation}
g=\Omega ^{2}=\epsilon =1,  \label{one}
\end{equation}%
while $m$ and $\gamma $ are kept as free parameters (a different
normalization, with $\epsilon \neq 1$, is adopted below in Section IV, where
a different potential is considered).

Equations (\ref{CGL1}) and (\ref{CGL2}) conserve the total norm,%
\begin{equation}
N=\int \int \left( \left\vert \phi _{1}\right\vert ^{2}+\left\vert \phi
_{2}\right\vert ^{2}\right) dxdy,  \label{N}
\end{equation}%
Hamiltonian,%
\begin{eqnarray}
H &=&\int \int \left[ \frac{1}{2m}\left( \left\vert \nabla \phi
_{1}\right\vert ^{2}+\left\vert \nabla \phi _{2}\right\vert ^{2}\right) +%
\frac{1}{2}\left( \left\vert \phi _{1}\right\vert ^{4}+\left\vert \phi
_{2}\right\vert ^{4}\right) +\gamma \left\vert \phi _{1}\right\vert
^{2}\left\vert \phi _{2}\right\vert ^{2}\right.  \notag \\
&&\left. +\frac{1}{2}\left( x^{2}+y^{2}\right) \left( \left\vert \phi
_{1}\right\vert ^{2}+\left\vert \phi _{2}\right\vert ^{2}-\left( \phi
_{1}\phi _{2}^{\ast }+\phi _{1}^{\ast }\phi _{2}\right) \right) \right] dxdy,
\label{H}
\end{eqnarray}%
where normalization (\ref{one}) is taken into regard, and $\ast $ stands for
the complex conjugate, as well as the total angular momentum,%
\begin{equation}
M=i\sum_{j=1,2}\int \int \phi _{j}^{\ast }\left( y\frac{\partial }{\partial x%
}-x\frac{\partial }{\partial y}\right) \phi _{j}dxdy.  \label{M}
\end{equation}

In terms of optics, Eqs. (\ref{CGL1}) and (\ref{CGL2}) with $t$ replaced by $%
z$ and $m$ replaced by the Fresnel number provide a model of the
spatial-domain copropagation of waves with orthogonal circular polarizations
in a bulk waveguide made of a self-defocusing material, provided that
relation $g=\gamma /2$ is imposed, which corresponds to the copropagation of
orthogonal circular polarizations. In that case, the HO potential represents
the transverse waveguiding structure, and the linear-coupling terms
represent linear mixing induced by anisotropy (linear birefringence) of the
material.

In Section II, we address the SSB effects in the GS produced by the full 2D
version of Eqs. \ref{CGL1}), (\ref{CGL2}) and their 1D reduction, under the
above-mentioned condition $\gamma >g$. The relevant solutions are obtained
in an analytical form by means of the Thomas-Fermi approximation (TFA), and
confirmed by full numerical solutions. Section III addresses vortex states,
with topological charges $S=1$ and $2$, in the 2D system, in which SSB is
again considered by means of TFA\ and a full numerical solution. In Section
IV we address the symmetry breaking of two-component gap solitons in the 1D
system, which is described by Eqs. (\ref{CGL1}) and (\ref{CGL2}) with the HO
potential replaced by the spatially periodic one, see Eqs. (\ref{gap1}) and (%
\ref{gap2}) below. In that case, the situation is drastically different, as
SSB takes place in \textit{antisymmetric} two-component solitons, under the
condition opposite to the one mentioned above, i.e., $\gamma <g$, which
means that the cross-repulsion is weaker than the intrinsic self-repulsion
The paper is concluded by Section V.

\section{Spontaneous symmetry breaking (SSB) of the ground states (GSs)}

Eigenstates of the system based on Eqs.~(\ref{CGL1}), (\ref{CGL2}), and (\ref%
{one}), with chemical potential $\mu >0$, are looked for as%
\begin{equation}
\phi _{1,2}\left( x,y\right) =\exp \left( -i\mu t\right) \varphi _{1,2}(x,y),
\label{varphi}
\end{equation}%
where functions $\varphi _{1,2}\left( x,y\right) $ satisfy stationary
equations
\begin{gather}
\mu \varphi _{1}=-\frac{1}{2m}\nabla ^{2}\varphi _{1}+|\varphi
_{1}|^{2}\varphi _{1}+\gamma \left\vert \varphi _{2}\right\vert ^{2}\varphi
_{1}+\frac{1}{2}\left( x^{2}+y^{2}\right) \varphi _{1}-\varphi _{2},
\label{1phi} \\
\mu \varphi _{2}=-\frac{1}{2m}\nabla ^{2}\varphi _{2}+|\varphi
_{2}|^{2}\varphi _{2}+\gamma \left\vert \varphi _{1}\right\vert ^{2}\varphi
_{2}+\frac{1}{2}\left( x^{2}+y^{2}\right) \varphi _{2}-\varphi _{1}.
\label{2phi}
\end{gather}%
The degree of asymmetry of solutions produced by Eqs. (\ref{1phi}) and (\ref%
{2phi}) is quantified by parameter%
\begin{equation}
R=\frac{2}{N}\int \int |\varphi _{1}\left( x,y\right) |^{2}dxdy-1,  \label{R}
\end{equation}%
where $N$ is the total norm defined as per Eq. (\ref{N}). Obviously, $R=0$
in the case when norms of both components are equal.

In the 1D case, $\nabla ^{2}$ is replaced by $\partial ^{2}/\partial x^{2}$,
$y^{2}$ is dropped in the HO potential, and functions $\varphi _{1,2}(x)$
are always real. In this case, the asymmetry is defined by the 1D version of
Eq. (\ref{R}), definitions of $N$ and $H$, given by Eqs. (\ref{N}) and (\ref%
{H}) are also replaced by their 1D versions, while the definition of the
angular momentum [see Eq. (\ref{M})] is irrelevant.

\subsection{The Thomas-Fermi approximation (TFA)}

First, we apply the TFA to Eqs. (\ref{1phi}) and (\ref{2phi}), dropping, as
usual, the derivatives in these equations \cite{BEC1,BEC2} (which is,
formally, tantamount to taking $m\rightarrow \infty $):
\begin{gather}
\mu \varphi _{1}=g\varphi _{1}^{3}+\gamma \varphi _{2}^{2}\varphi _{1}+\frac{%
1}{2}\left( x^{2}+y^{2}\right) \varphi _{1}-\varphi _{2},  \label{1} \\
\mu \varphi _{2}=g\varphi _{2}^{3}+\gamma \varphi _{1}^{2}\varphi _{2}+\frac{%
1}{2}\left( x^{2}+y^{2}\right) \varphi _{2}-\varphi _{1}.  \label{2}
\end{gather}%
Equations (\ref{1}) and (\ref{2}) admit two nonzero solutions: an obvious
\textit{symmetric} one,%
\begin{equation}
\varphi _{1}^{2}(x,y)=\varphi _{2}^{2}(x,y)=\frac{2\left( \mu +1\right)
-\left( x^{2}+y^{2}\right) }{2\left( \gamma +1\right) },  \label{symm}
\end{equation}%
which exists at
\begin{equation}
\frac{1}{2}\left( x^{2}+y^{2}\right) <\mu +1,  \label{U<<}
\end{equation}%
and a \textit{symmetry-broken} (alias asymmetric)\textit{\ }solution, which
exists under the condition adopted above, $\gamma >1$ (the cross-repulsion
is stronger than self-repulsion):
\begin{equation}
\varphi _{1}^{2}+\varphi _{2}^{2}=\mu -\frac{1}{2}\left( x^{2}+y^{2}\right)
,~\varphi _{1}\varphi _{2}=\frac{1}{\gamma -1}.  \label{asymm}
\end{equation}%
Formally, solution (\ref{asymm}) exists at $\gamma <1$ too, but, having $%
\varphi _{1}\varphi _{2}<0$, it will make the last (linear-coupling) term in
Hamiltonian (\ref{H}) positive, which definitely implies instability of the
respective states \cite{Peng-book}.

The asymmetric solution given by Eq. (\ref{asymm}) exists in a region where
it complies with obvious condition $\varphi _{1}^{2}+\varphi
_{2}^{2}>2\varphi _{1}\varphi _{2}$, i.e.,%
\begin{equation}
x^{2}+y^{2}<2\left( \mu -\frac{2}{\gamma -1}\right) .  \label{U<}
\end{equation}%
Note that condition (\ref{U<}) may hold if, at least, it is valid at $x=y=0$%
, thus a condition necessary for the existence of the asymmetric solution is%
\begin{equation}
\mu >\mu ^{\mathrm{(cr)}}\equiv \frac{2}{\gamma -1}.  \label{mu}
\end{equation}%
Thus, TFA produces the state with the spontaneously broken symmetry which
features a \textit{two-layer structure}:%
\begin{equation}
\left\{
\begin{array}{c}
\mathrm{asymmetric,~given~by~Eq.(\ref{asymm}),~in~the~}inner~(central)~layer%
\mathrm{,}~ \\
0\leq x^{2}+y^{2}<2\left( \mu -2\left( \gamma -1\right) ^{-1}\right) ; \\
\mathrm{symmetric,~given~by~Eq.(\ref{symm}),~in~the~}%
outer~(surrounding)~layer\mathrm{,}~ \\
2\left( \mu -2\left( \gamma -1\right) ^{-1}\right) \leq x^{2}+y^{2}<2\left(
\mu +1\right) \equiv r_{\mathrm{outer}}^{2};\mathrm{~}%
\end{array}%
\right.   \label{2-layer}
\end{equation}%
and zero outside of both layers, i.e., at $x^{2}+y^{2}>r_{\mathrm{outer}}^{2}
$. Of course, the exact solution has a small nonzero \textquotedblleft tail"
in the latter area, which is, as usual, ignored by TFA. Another difference
of a numerically exact solution is that the symmetry is slightly broken in
the TFA-symmetric layer.

If condition (\ref{mu}) does not hold, the inner layer does not exist, and
the entire TFA solution keeps the symmetric form, as given by Eqs. (\ref%
{symm}) and (\ref{U<<}). Its total norm, defined as per Eq. (\ref{N}) or its
1D version, is%
\begin{eqnarray}
\left( N_{\mathrm{TFA}}^{\mathrm{(symm)}}\right) _{\mathrm{2D}} &=&2\pi
\frac{\left( \mu +1\right) ^{2}}{\gamma +1},  \label{N2D} \\
\left( N_{\mathrm{TFA}}^{\mathrm{(symm)}}\right) _{\mathrm{1D}} &=&\frac{%
2^{7/2}}{3}\frac{\left( \mu +1\right) ^{3/2}}{\gamma +1},  \label{N1D}
\end{eqnarray}%
With the increase of the cross-repulsion strength, $\gamma $, at a fixed
norm, the SSB sets in when $\mu $, expressed in terms of the norm of the
symmetric GS by means of Eq. (\ref{N1D}) or (\ref{N2D}), attains critical
value (\ref{mu}). After a simple algebra, this condition leads to equations
which predict, in the framework of TFA, the critical value of $\gamma $,
above which SSB takes place for given $N$,
\begin{gather}
\frac{3N}{2^{7/2}}=\frac{\sqrt{\gamma _{\mathrm{1D}}^{\mathrm{(cr)}}+1}}{%
\left( \gamma _{\mathrm{1D}}^{\mathrm{(cr)}}-1\right) ^{3/2}},  \label{1D} \\
\frac{N}{2\pi }=\frac{\gamma _{\mathrm{2D}}^{\mathrm{(cr)}}+1}{\left( \gamma
_{\mathrm{2D}}^{\mathrm{(cr)}}-1\right) ^{2}}.  \label{2D}
\end{gather}%
In particular, for $N$ large enough, when TFA is a natural approximation,
Eqs. (\ref{1D}) and (\ref{2D}) yield critical values of $\gamma $ close to $%
g\equiv 1$, \textit{viz}., $\gamma _{\mathrm{1D}}^{\mathrm{(cr)}}\approx
1+\left( 16/3N\right) ^{2/3}$, $\gamma _{\mathrm{2D}}^{\mathrm{(cr)}}\approx
1+2\sqrt{\pi /N}$.

\subsection{Numerical results for 1D states}

\begin{figure}[h]
\begin{center}
\includegraphics[height=4.cm]{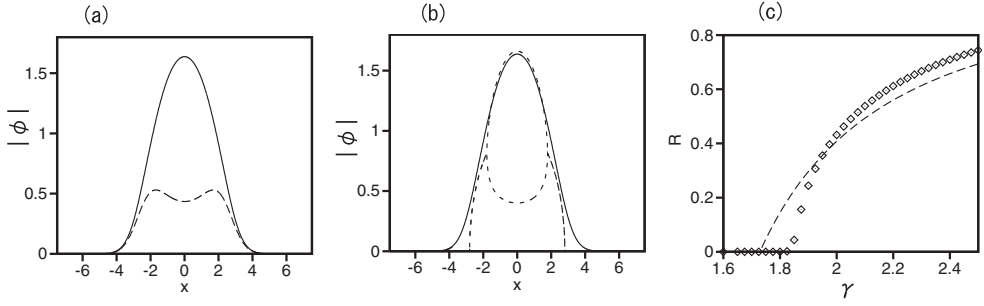}
\end{center}
\caption{(a) Profiles of components $\left\vert \protect\phi %
_{1}(x)\right\vert $ ans $\left\vert \protect\phi _{2}(x)\right\vert $
(solid and dashed lines) in the stable 1D ground state (GS) with broken
symmetry between the components. The solution was obtained by means of the
imaginary-time-evolution method, applied to the 1D version of Eqs. (\protect
\ref{CGL1}) and (\protect\ref{CGL2}) with $\protect\gamma =2.5$ and $m=1$,
for the total norm $N=10$ [see Eq. (\protect\ref{N})]. (b) Comparison of $%
\left\vert \protect\phi _{1}(x)\right\vert $ (the solid line) taken from the
same numerical solution, and its TFA counterpart, produced by Eqs. (\protect
\ref{symm}), (\protect\ref{asymm}), and (\protect\ref{2-layer}) (the dashed
line running close to the solid one); the TFA-produced component $\left\vert
\protect\phi _{2}(x)\right\vert $ is shown too, by the double-peaked dashed
curve. (c) The asymmetry measure, defined as per Eq. (\protect\ref{R}), vs. $%
\protect\gamma $, for fixed $m=10$ and $N=10$. The chain of rhombuses and
the dashed line show, severally, the numerical results and their
TFA-produced counterparts.}
\label{fig1}
\end{figure}

Numerical solutions for the 1D version of Eqs. (\ref{CGL1}), (\ref{CGL2})
and (\ref{1phi}), (\ref{2phi}) are presented here for the total norm $N=10$,
as defined by the 1D form of Eq. (\ref{N}), since this value makes it
possible to produce generic results. Figure \ref{fig1}(a) displays profiles
of components $\varphi _{1,2}(x)\equiv \left\vert \phi _{1,2}(x)\right\vert $
of the GS, obtained by means of the imaginary-time evolution method \cite%
{IT,IT2} applied to Eqs. (\ref{CGL1}), (\ref{CGL2}), for $\gamma =2.5$ and $%
m=1$. Further, Fig. \ref{fig1}(b) produces TFA profiles for the same
parameters, constructed as per Eqs. (\ref{symm}), (\ref{asymm}), and (\ref%
{2-layer}), and provides explicit comparison of this approximate analytical
solution for $\varphi _{1}(x)$ with its numerically found counterpart [the
comparison for component $\varphi _{2}(x)$ is provided by the juxtaposition
of panels (a) and (b) in Fig. \ref{fig1}].

Figure \ref{fig1}(c) summarizes the numerical and analytical results by
plotting the asymmetry degree $R$, defined as per Eq. (\ref{R}), vs. the
cross-repulsion strength, $\gamma $, at a fixed value of the norm, $N=10$,
and fixed mass, $m=1$, as obtained from the numerical solution, and from
TFA, i.e., as produced by the integration of expressions (\ref{symm}), (\ref%
{asymm}), and (\ref{2-layer}). It is seen that SSB takes place at $\gamma
>\gamma _{\mathrm{numer}}^{\mathrm{(cr)}}\approx 1.83$, while its TFA
counterpart, obtained from Eq. (\ref{1D}), is $\gamma _{\mathrm{TFA}}^{%
\mathrm{(cr)}}\approx 1.73$ for $N=10$. The analytically predicted curve $%
R(\gamma )$ is reasonably close to its numerical counterpart, both showing
the SSB transition which may be identified as a \textit{supercritical
bifurcation} \cite{bif}.

\begin{figure}[h]
\begin{center}
\includegraphics[height=4.cm]{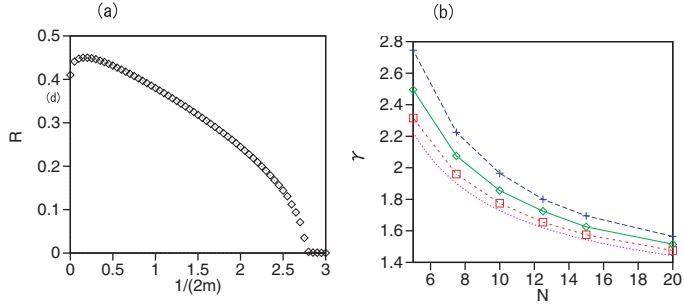}
\end{center}
\caption{(a) The numerically found asymmetry measure $R$ for the 1D GS
solutions [see Eq. (\protect\ref{R})], as a function of the inverse-mass
parameter, $1/(2m)$. (b) The critical values of $\protect\gamma $, above
which the GS symmetry is broken in 1D, as a function of $N$ for fixed values
of the inverse-mass coefficient, $1/(2m)=2$ (pluses), $0.5$ (rhombuses), and
$0.05$ (squares). The dotted curve is the analytical result predicted by
TFA, as per Eq. (\protect\ref{1D}). }
\label{fig2}
\end{figure}

Overall, TFA provides reasonable accuracy for $m=1$, even if, formally
speaking, this approximation applies for small values of coefficient $1/(2m)$
in Eqs. (\ref{1phi}) and (\ref{2phi}). To focus on the role of this
parameter, Fig. \ref{fig2} displays $R$ vs. $1/(2m)$ at $\gamma =2.5$. The
SSB disappears at $1/(2m)>2.8$. Another essential characteristic of the GS
solutions is the critical value of $\gamma $, above which SSB sets in;
recall that, in the framework of TFA, it is predicted by Eq. (\ref{1D}). A
set of curves of $\gamma _{\mathrm{1D}}^{\mathrm{(cr)}}(N)$ for several
fixed values of the inverse-mass parameter, \textit{viz}., $1/\left(
2m\right) =2$, $0.5$, and $0.05$, are plotted in Fig. \ref{fig2}(b), along
with the corresponding TFA limit, corresponding to $1/\left( 2m\right) =0$.

\begin{figure}[h]
\begin{center}
\includegraphics[height=4.cm]{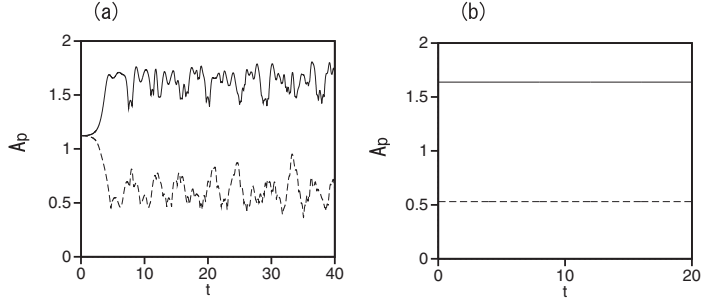}
\end{center}
\caption{The evolution of the largest values of $|\protect\phi _{1}(x)|$ and
$\left\vert \protect\phi _{2}(x)\right\vert $ (solid and dashed lines),
produced by direct simulations of the 1D version of Eqs.~(\protect\ref{CGL1}%
) and (\protect\ref{CGL2}) for $\protect\gamma =2.5$, $m=1$, and total norm $%
N=10$, starting from (a) the unstable symmetrical state and (b) the stable
GS with broken symmetry.}
\label{fig3}
\end{figure}

The SSB effect corresponding to the supercritical bifurcation implies that,
when symmetric and symmetry-broken stationary solutions coexist as
stationary ones, the symmetric solution should be unstable. This expectation
is corroborated in Fig. \ref{fig3}(a), which shows the time evolution of the
maximum values of $|\phi _{1}(x)|$ and $|\phi _{2}(x)|$, produced by direct
simulations of coupled equations~(\ref{CGL1}) and (\ref{CGL2}), for $\gamma
=2.5$, $m=1$, and $N=10$. The input is a symmetric solution, with $\varphi
_{1}(x)=\varphi _{2}(x)$, and a small perturbation added to it. It is
unstable, clearly tending to spontaneously transform into a broken-symmetry
state. On the other hand, Fig. \ref{fig3}(b) demonstrates stability of the
asymmetric GS for the same values of the parameters.

\section{Symmetry-breaking 2D vortex states}

\subsection{The Thomas-Fermi approximation (TFA)}

Solutions of the system of coupled 2D equations (\ref{CGL1}) and (\ref{CGL2}%
) with embedded angular momentum are looked for, in the polar coordinates $%
\left( r,\theta \right) $, as
\begin{equation}
\phi _{1,2}=\exp \left( -i\mu t+iS\theta \right) \varphi _{1,2}(r),
\label{S}
\end{equation}%
where $\varphi _{1,2}(r)$ are real radial wave functions, and $S=1,2,3,...$
is the integer vorticity, $S=0$ corresponding to the 2D GS. For stationary
states represented by ansatz (\ref{S}), the angular momentum, defined as per
Eq. (\ref{M}), is related to the total norm, $M=SN$.

In the framework of TFA (which was previously applied to delocalized vortex
states \cite{Fetter}), the substitution of ansatz (\ref{S}) leads to the
following radial equations, instead of 1D equations (\ref{1}) and (\ref{2}):
\begin{gather}
\mu \varphi _{1}=g\varphi _{1}^{3}+\gamma \varphi _{2}^{2}\varphi _{1}+\frac{%
1}{2}\left( r^{2}+\frac{S^{2}}{r^{2}}\right) \varphi _{1}-\varphi _{2},
\label{S1} \\
\mu \varphi _{2}=g\varphi _{2}^{3}+\gamma \varphi _{1}^{2}\varphi _{2}+\frac{%
1}{2}\left( r^{2}+\frac{S^{2}}{r^{2}}\right) \varphi _{2}-\varphi _{1}.
\label{S2}
\end{gather}%
Straightforward consideration of Eqs. (\ref{S1}) and (\ref{S2}) demonstrates
that solutions are different from zero in annulus%
\begin{equation}
r_{\mathrm{core}}^{2}<r^{2}<r_{\mathrm{outer}}^{2},  \label{core-outer}
\end{equation}%
\begin{equation}
r_{\mathrm{outer,core}}^{2}=\mu +1\pm \sqrt{\left( \mu +1\right) ^{2}-S^{2}},
\label{outer-core}
\end{equation}%
provided that $\mu +1>S$. Note that the central \textquotedblleft empty"
core, in which the TFA solution is zero, is absent in the GS solution (with $%
S=0$), as Eq. (\ref{outer-core}) yields $r_{\mathrm{core}}^{2}\left(
S=0\right) =0$. In the same case, Eq. (\ref{outer-core}) yields $r_{\mathrm{%
outer}}^{2}\left( S=0\right) =2\left( \mu +1\right) $, which coincides with $%
r_{\mathrm{outer}}^{2}$ given above by Eq. (\ref{2-layer}) for the GS.

In annulus (\ref{core-outer}), the symmetric solution of Eqs. (\ref{S1}) and
(\ref{S2}) is%
\begin{equation}
\varphi _{1,2}^{2}(r)\equiv \varphi ^{2}(r)=\frac{\mu +1}{\gamma +1}-\frac{1%
}{2\left( \gamma +1\right) }\left( r^{2}+\frac{S^{2}}{r^{2}}\right) .
\label{symm2D}
\end{equation}%
Note that setting $S=0$ makes this solution identical to its GS counterpart
given by Eq. (\ref{symm}). The total norm of symmetric expression (\ref%
{symm2D}) is%
\begin{gather}
N_{\mathrm{TFA}}^{\mathrm{(symm)}}(S)=2\pi \frac{\mu +1}{\gamma +1}\sqrt{%
\left( \mu +1\right) ^{2}-S^{2}}  \notag \\
-\frac{2\pi S^{2}}{\gamma +1}\ln \left( \frac{\mu +1+\sqrt{\left( \mu
+1\right) ^{2}-S^{2}}}{S}\right) ,  \label{N(S)}
\end{gather}%
which, for $S=0$ (the GS), is tantamount to Eq. (\ref{N2D}).

Points at which the broken-symmetry solution with $S\geq 1$ (if any)
branches off from the symmetric one, given by Eq. (\ref{symm2D}), are
determined by taking the difference of Eqs. (\ref{S1}) and (\ref{S2}),
linearizing it for $\varphi _{1}-\varphi _{2}\rightarrow 0$, and cancelling
the common infinitesimal factor $\left( \varphi _{1}-\varphi _{2}\right) $:%
\begin{equation}
\mu -1 =\left( 3-\gamma \right) \varphi ^{2}(r)+\frac{1}{2}\left(
r^{2}+\frac{S^{2}}{r^{2}}\right) .  \label{linearized}
\end{equation}%
The substitution of the symmetric TFA solution (\ref{symm2D}) in Eq. (\ref%
{linearized}) leads to the conclusion that the asymmetric solution may exist
between the following branching points:%
\begin{equation}
r_{\max ,\min }^{2}=\mu -\frac{2}{\gamma -1}\pm \sqrt{\left( \mu -\frac{2}{%
\gamma -1}\right) ^{2}-S^{2}}  \label{maxmin}
\end{equation}

Thus, TFA predicts the vortex solution with a three-layer structure. First,
it vanishes in the empty core and in the peripheral zone,
\begin{equation}
\varphi _{1,2}\left( r<r_{\mathrm{core}}\right) =\varphi _{1,2}\left( r>r_{%
\mathrm{outer}}\right) =0.  \label{core}
\end{equation}%
The symmetric solution, as given by Eq. (\ref{symm2D}), is supported in two
edge layers,
\begin{eqnarray}
r_{\mathrm{core}}^{2} &<&r^{2}<r_{\min }^{2},  \label{r0rmin} \\
r_{\max }^{2} &<&r^{2}<r_{\mathrm{outer}}^{2}~,  \label{rmaxrouter}
\end{eqnarray}%
with the edges determined by Eqs. (\ref{outer-core}) and (\ref{maxmin}).
Finally, the broken symmetry is featured by the TFA solution in the inner
layer,
\begin{equation}
r_{\min }^{2}<r^{2}<r_{\max }^{2}.  \label{rminrmax}
\end{equation}

In the framework of TFA, the condition necessary for the existence of the
asymmetric solution amount to the existence of real values (\ref{maxmin}),
i.e.,%
\begin{equation}
\mu >\mu ^{\mathrm{(cr)}}(S)\equiv \frac{2}{\gamma -g}+S,  \label{mu2D}
\end{equation}%
cf. Eq. (\ref{mu}) for the GS. Further, combining Eq. (\ref{mu2D}) with
expression (\ref{N(S)}) for the norm of the symmetric vortex state, one can
derive an equation for the critical strength of the cross-repulsion, above
which the symmetry of the vortex states is broken, cf. Eq. (\ref{2D}) for
the 2D GS ($S=0$). In particular, for $S=1$ the equation is
\begin{gather}
N=\frac{2\pi \gamma _{S=1}^{\mathrm{(cr)}}}{\left( \gamma _{S=1}^{\mathrm{%
(cr)}}-1\right) ^{2}}\sqrt{\frac{3\gamma _{S=1}^{\mathrm{(cr)}}-1}{\gamma
_{S=1}^{\mathrm{(cr)}}+1}}  \notag \\
-\frac{2\pi }{\gamma _{S=1}^{\mathrm{(cr)}}+1}\ln \left( \frac{2\gamma
_{S=1}^{\mathrm{(cr)}}+\sqrt{\left( \gamma _{S=1}^{\mathrm{(cr)}}+1\right)
\left( 3\gamma _{S=1}^{\mathrm{(cr)}}-1\right) }}{\gamma _{S=1}^{\mathrm{(cr)%
}}-1}\right) .  \label{S=1}
\end{gather}

\subsection{Numerical results for 2D ground and vortex states}

\begin{figure}[h]
\begin{center}
\includegraphics[height=3.5cm]{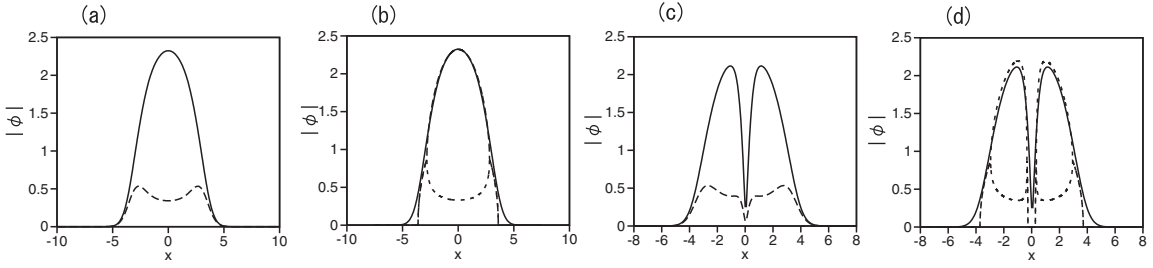}
\end{center}
\caption{(a) 1D cross sections (radial profiles) of components $\left\vert
\protect\phi _{1}\right\vert $ ans $\left\vert \protect\phi _{2}\right\vert $
(solid and dashed lines, respectively) of the stable 2D ground state ($S=0$)
with broken inter-component symmetry. The solution was obtained by means of
the imaginary-time-evolution method applied to the axisymmetric reduction of
Eqs. (\protect\ref{CGL1}) and (\protect\ref{CGL2}) in the 2D form, with $%
\protect\gamma =2.3$ and $m=1$, for the total 2D norm $N=100$. (b)
Comparison of $\left\vert \protect\phi _{1}\right\vert $ (the solid line)
taken from the same numerical solution as in (a), and its TFA counterpart,
produced by Eqs. (\protect\ref{S1}) and (\protect\ref{S2}) (the dashed line
which almost completely overlaps with the solid one). The other dashed line
shows the TFA-produced component $\left\vert \protect\phi _{2}\right\vert $.
(c,d): The same as in (a,b), but for a stable vortex mode with $S=1$ and the
same values of $\protect\gamma $, $m$, and $N$.}
\label{fig4}
\end{figure}

Proceeding to numerical results for the 2D system, in Fig. \ref{fig4}(a) we,
first, display numerically found radial profiles, $\varphi _{1,2}(r)\equiv
\left\vert \phi _{1,2}(r)\right\vert $, of the two components of the 2D GS
(with $S=0$), for $\gamma =2.3$, $m=1$, and the total 2D norm $N=100$. The
numerical profiles are compared to their TFA-predicted counterparts,
obtained from Eqs. (\ref{S1}) and (\ref{S2}), in Fig. \ref{fig4}(b).
Further, the numerical and approximate analytical radial profiles of a
stable vortex state, with $S=1$ and the same values of other parameters as
in panels (a,b), are displayed in Figs. \ref{fig4}(c,d). In addition to the
2D states with $S=0$ and $S=1$, higher-order ones, with $S\geq 2$ have been
constructed too (not shown here in detail).

\begin{figure}[h]
\begin{center}
\includegraphics[height=4.cm]{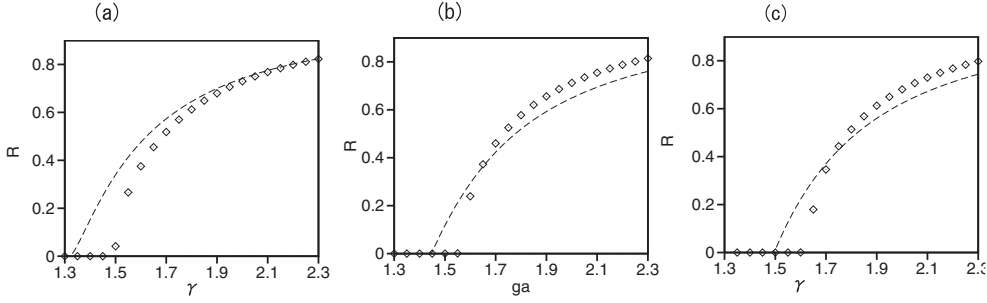}
\end{center}
\caption{The asymmetry measure (\protect\ref{R}) vs. the cross-repulsion
strength, $\protect\gamma $, for families of 2D states with $S=0$ [the
ground state, (a)], $S=1$ (b), and $S=2$ (c), with fixed effective mass, $m=1
$, and 2D norm, $N=100$. Chains of rhombuses and dashed curves represent,
severally, numerical results produced by the imaginary-time-propagation
method, and analytical results provided by TFA, \textit{viz}., Eqs. (\protect
\ref{asymm}) and (\protect\ref{2-layer}) for $S=0$, and Eqs. (\protect\ref%
{S1}), (\protect\ref{S2}) for $S=1$ and $2$.}
\label{fig5}
\end{figure}

The results are summarized in Figs. \ref{fig5}(a), (b), and (c) by means of
curves $R(\gamma )$ for the dependence of asymmetry measure (\ref{R}) on the
cross-component repulsion strength, $\gamma $, for the fixed mass, $m=1$,
total norm, $N=100$, and three values of the vorticity, $S=0$ [(a), the GS],
$S=1$ (b), and $S=2$ (c). The TFA-predicted analytical results are included
too. In particular, for $S=1$ the numerically found SSB point is $\gamma
_{S=1}^{\mathrm{(cr)}}\approx 1.55$, while its TFA counterpart, found from
Eq. (\ref{S=1}), is $\left( \gamma _{S=1}^{\mathrm{(cr)}}\right) _{\mathrm{%
TFA}}\approx 1.45$. Similar to the conclusion made above for 1D GS [cf. Fig. %
\ref{fig1}(c)], for all the cases of $S=0$, $1$, and $2$ the SSB
transitions, displayed in Figs. \ref{fig5}(a-c), may be categorized as the
supercritical bifurcation \cite{bif}.

In comparison with the similar results for the 1D\ GS, displayed in Fig. \ref%
{fig1}(c), the relative error of TFA [in particular, in predicting $\gamma ^{%
\mathrm{(cr)}}$] is much larger in the 2D setting, being $\simeq 5\%$ in 1D
and $\simeq 15\%$ in 2D. On the other hand, it is worthy to note that the
relative error is smaller by a factor $\simeq 3$ for the vortex states with $%
S=1$ and $2$, in comparison to the 2D GS.

\begin{figure}[h]
\begin{center}
\includegraphics[height=4.cm]{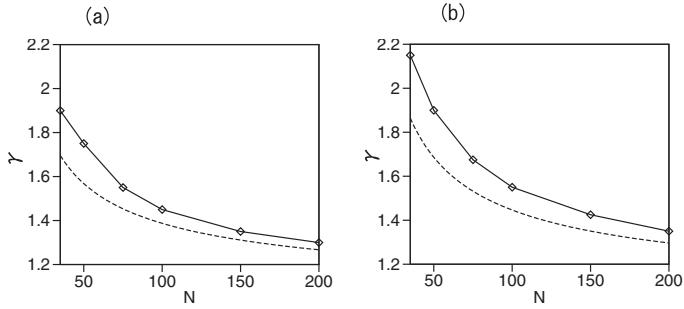}
\end{center}
\caption{The critical value of the cross-repulsion strength at the
symmetry-breaking point in the 2D system, as a function of $N$ at $m=1$. (a)
The ground state, $S=0$; (b) the vortex, with $S=1$. The dashed lines denote
the TFA prediction, produced, respectively, by Eqs. (\protect\ref{2D}) and (%
\protect\ref{S=1}).}
\label{fig6}
\end{figure}

Numerical and approximate analytical dependences of the value of $\gamma $
at the SSB point, $\gamma ^{\mathrm{(cr)}}$, on the total norm, $N$, for the
2D GS ($S=0$) and vortex mode with $S=1$ are plotted in Fig. \ref{fig6},
fixing the effective mass to be $m=1$. The analytical curves are produced by
TFA, i.e., Eq. (\ref{2D}) for $S=0$, and Eq. (\ref{S=1}) for $S=1$. In
comparison to similar results for the 1D\ GS, shown in Fig. \ref{fig2}(b),
the accuracy of TFA in 2D is only slightly poorer (note, however, a great
difference in the scale of $N$ between the 1D and 2D cases).

\begin{figure}[h]
\begin{center}
\includegraphics[height=4.cm]{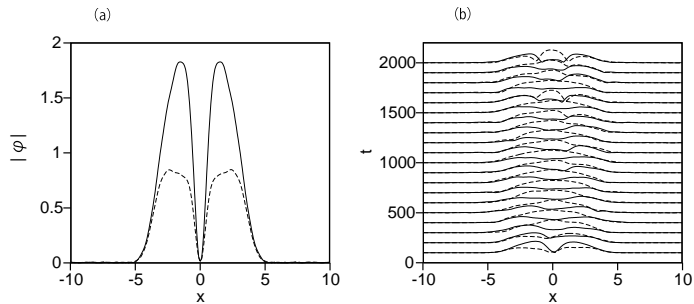}
\end{center}
\caption{(a) A numerically found 2D stationary state with double vorticity, $%
S=2$ and broken symmetry between the components. (b) Unstable evolution of
this state, leading to its spontaneous splitting in two unitary vortices.
The parameters are $m=1$ and $\protect\gamma =1.8$, the total norm of the
stationary state being $N=100$. Both the stationary state and the perturbed
evolution are displayed by means of 1D cross sections. }
\label{fig7}
\end{figure}

Finally, systematic direct simulations of the perturbed evolution of the 2D
states clearly demonstrate that families of the asymmetric GS solutions with
$S=0$ and unitary-vortex\ ones with $S=1$ are completely stable, while
asymmetric double vortices, with $S=2$, are not. A typical example of the
instability development in displayed in Fig. \ref{fig7}, which shows
spontaneous splitting of the double vortex into a pair of unitary ones, that
keep the asymmetric structure, with respect to components $\phi _{1}$ and $%
\phi _{2}$. It is relevant to mention that double vortices may be unstable
in the framework of the single GPE with the HO trapping potential and
self-repulsive nonlinearity, while unitary vortices are completely stable in
the same setting \cite{HPu}.

\section{Spontaneous symmetry breaking (SSB)\ in two-component gap solitons}

Another possibility to create localized states in the presence of a fully
repulsive nonlinearity, is offered, instead of the HO trapping potential, by
a periodic one -- namely, an optical lattice in BEC \cite{Oberthaler,Morsch}%
, or a photonic-crystal structure in optics \cite{PhotCryst1,PhotCryst2}. It
is well known that the interplay of the self-repulsion with a periodic
potential gives rise to stable gap solitons \cite{gap-sol,Lisbon}. The
analysis of two-component gap solitons in linearly-coupled dual-core systems
with a periodic potential and intrinsic self-repulsive nonlinearity in each
layer (in the absence of the inter-component nonlinearity) was developed in
Refs. \cite{Arik-1D,Arik,Marek2}. As mentioned above, symmetric states in
such systems are not subject to SSB, while antisymmetric ones develop
instability which replaces them by states with broken antisymmetry. Here, we
aim to demonstrate that SSB\ occurs in the coupled system if it includes
\textit{relatively weak} repulsion between the components (instead of the
relatively strong inter-component repulsion, which was necessary for the SSB
effect considered above in the states trapped in the HO potential).

In the 1D setting, the system of linearly coupled GPEs, including the
periodic potential with amplitude $U_{0}$, is written, in the scaled form,
as
\begin{gather}
i\frac{\partial \phi _{1}}{\partial t}=-\frac{1}{2}\frac{\partial ^{2}\phi
_{1}}{\partial x^{2}}-U_{0}\cos \left( 2\pi x\right) \phi _{1}+(|\phi
_{1}|^{2}+\gamma |\phi _{2}|^{2})\phi _{1}-\epsilon \phi _{2},  \label{gap1}
\\
i\frac{\partial \phi _{2}}{\partial t}=-\frac{1}{2}\frac{\partial ^{2}\phi
_{2}}{\partial x^{2}}-U_{0}\cos \left( 2\pi x\right) \phi _{2}+(|\phi
_{2}|^{2}+\gamma |\phi _{1}|^{2})\phi _{2}-\epsilon \phi _{1}.  \label{gap2}
\end{gather}%
Here, the period of the potential is fixed to be $1$, and, to clearly
demonstrate the SSB effects, it is convenient to fix the linear-coupling
constant as $\epsilon =0.05$, instead of $\epsilon =1$ in Eqs. (\ref{CGL1}),
(\ref{CGL2}).

To develop an analytical approach to the study of the gap solitons, we
resort to the averaging method \cite{Baizakov}, which looks for solutions in
the form of a rapidly oscillating carrier wave, with the period equal to the
double period of the lattice potential, modulated by slowly varying
(envelope) wave functions, $\Phi _{1,2}(x,t)$:%
\begin{equation}
\phi _{1,2}\left( x,t\right) =\Phi _{1,2}\left( x,t\right) \cos \left( \pi
x\right) .  \label{Phi}
\end{equation}%
It is known that this approach makes it possible to predict gap solitons
existing close to edges of the spectral bandgap. The substitution of this
ansatz in Eqs. (\ref{gap1}) and (\ref{gap2}) and application of the
averaging procedure leads, in the lowest approximation, to the following
system of equations for the slowly varying wave functions:
\begin{gather}
i\frac{\partial \Phi _{1}}{\partial t}=-\frac{1}{2m_{\mathrm{eff}}}\frac{%
\partial ^{2}\Phi _{1}}{\partial x^{2}}+\frac{3}{4}(|\Phi _{1}|^{2}+\gamma
|\Phi _{2}|^{2})\Phi _{1}-\epsilon \Phi _{2},  \label{gap3} \\
i\frac{\partial \Phi _{2}}{\partial t}=-\frac{1}{2m_{\mathrm{eff}}}\frac{%
\partial ^{2}\Phi _{2}}{\partial x^{2}}+\frac{3}{4}(|\Phi _{2}|^{2}+\gamma
|\Phi _{1}|^{2})\Phi _{2}-\epsilon \Phi _{1},  \label{gap4}
\end{gather}%
where the effective mass is
\begin{equation}
m_{\mathrm{eff}}=|U_{0}|/(|U_{0}|-2\pi ^{2}).  \label{meff}
\end{equation}%
At $U_{0}<2\pi ^{2}$, the effective mass is negative, hence its interplay
with the repulsive sign of the nonlinearity in Eqs. (\ref{gap3}) and (\ref%
{gap4}) gives rise to bright solitons, similar to how usual solitons are
produced by the balance of the positive mass and attractive nonlinearity.

To use the results reported in Ref. \cite{we1}, it is convenient to
additionally rescale the variables in Eqs. (\ref{gap3}) and (\ref{gap4}),
defining $t^{\prime }=-\epsilon t$, $x^{\prime }=\sqrt{-m_{\mathrm{eff}%
}\epsilon }x$, and
\begin{equation}
\Phi _{1}^{\prime }=\sqrt{4\epsilon /(3\gamma )}\Phi _{1},\Phi _{2}^{\prime
}=-\sqrt{4\epsilon /(3\gamma )}\Phi _{2},  \label{-}
\end{equation}%
thus replacing Eqs. (\ref{gap3}) and (\ref{gap4}) by
\begin{gather}
i\frac{\partial \Phi _{1}^{\prime }}{\partial t^{\prime }}=-\frac{1}{2}\frac{%
\partial ^{2}\Phi _{1}^{\prime }}{\partial x^{\prime 2}}-\left( \frac{1}{%
\gamma }|\Phi _{1}^{\prime }|^{2}+|\Phi _{2}^{\prime }|^{2}\right) \Phi
_{1}^{\prime }-\Phi _{2}^{\prime }.  \label{gap5} \\
i\frac{\partial \Phi _{2}^{\prime }}{\partial t^{\prime }}=-\frac{1}{2}\frac{%
\partial ^{2}\Phi _{2}^{\prime }}{\partial x^{\prime 2}}-\left( \frac{1}{%
\gamma }|\Phi _{2}^{\prime }|^{2}+|\Phi _{1}^{\prime }|^{2}\right) \Phi
_{2}^{\prime }-\Phi _{1}^{\prime }.  \label{gap6}
\end{gather}%
Accordingly, the total norm of rescaled fields $\Phi _{1,2}^{\prime }$ is
related to the norm of the original ones, $\Phi _{1,2}$:%
\begin{equation}
N^{\prime }=(3\gamma /4)\sqrt{m_{\mathrm{eff}}/\epsilon }N.  \label{N'}
\end{equation}

As shown in Ref. \cite{we1}, Eqs. (\ref{gap5}) and (\ref{gap6}) with $\gamma
<1$ give rise to SSB of symmetric solitons, in terms of this system, i.e.,
ones with $\Phi _{1}^{\prime }=\Phi _{2}^{\prime }$, if $\gamma $ belongs to
interval $\gamma <\left( \gamma ^{\prime }\right) ^{\mathrm{(cr)}}$, where
the respective critical value is related to $N^{\prime }$ by equation
\begin{equation}
N^{\prime }=\frac{256\left[ 1/\left( \gamma ^{\prime }\right) ^{\mathrm{(cr)}%
}+1\right] ^{-1}}{\left[ \sqrt{25/\left( \gamma ^{\prime }\right) ^{\mathrm{%
(cr)}}-7}-3\sqrt{1/\left( \gamma ^{\prime }\right) ^{\mathrm{(cr)}}+1}\right]
\left[ \sqrt{25/\left( \gamma ^{\prime }\right) ^{\mathrm{(cr)}}-7}+\sqrt{%
1/\left( \gamma ^{\prime }\right) ^{\mathrm{(cr)}}+1}\right] }.  \label{NC}
\end{equation}%
Actually, Eq. (\ref{-}) implies that the solitons subject to constraint $%
\Phi _{1}^{\prime }=\Phi _{2}^{\prime }$ represent \textit{antisymmetric}
solitons, with $\Phi _{2}(x)=-\Phi _{1}(x)$, in terms of the original
equations (\ref{gap3}) and (\ref{gap4}).

\begin{figure}[h]
\begin{center}
\includegraphics[height=4.cm]{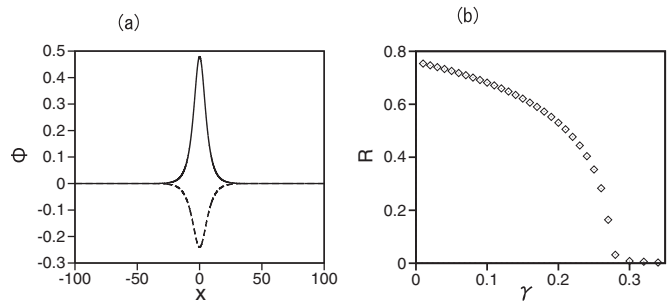}
\end{center}
\caption{(a) Profiles of $\Phi _{1}$ and $\Phi _{2}$ (solid and dashed
lines, respectively), obtained as a numerical solution to Eqs.~(\protect\ref%
{gap3}) and (\protect\ref{gap4}), representing the two-component soliton
with broken antisymmetry, for $U_{0}=5$, $\protect\epsilon =0.05$, and total
norm $N=2.5$. (b) The asymmetry measure $R(\protect\gamma )$, defined for
the family of the two-component solitons, as produced by the numerical
solution for $U_{0}=5$, $\protect\epsilon =0.05$, and total norm $N=2.5$. }
\label{fig8}
\end{figure}

Numerical solutions for asymmetric solitons can be readily generated by the
imaginary-time integration method applied to Eqs. (\ref{gap3}) and (\ref%
{gap4}). Figure \ref{fig8}(a) shows an example of a soliton with broken
antisymmetry, obtained at parameters $\gamma =0.2$, $U_{0}=5$, $\epsilon
=0.05$, and $N=2.5$. Note that the signs of the two components are opposite,
in accordance with what is said above, and the breaking of the antisymmetry
is exhibited by the difference of their amplitudes. Further, Fig. \ref{fig8}%
(b) summarizes the results, by displaying the asymmetry measure $R$ for the
soliton family, defined as per Eq. (\ref{R}) with $\left\vert \varphi
_{1}\right\vert ^{2}$ replaced by $\left\vert \Phi _{1}\right\vert ^{2}$,
and $N$ replaced by the total norm defined in terms of envelope wave
functions $\Phi _{1,2}$, versus the relative strength $\gamma $ of the
inter-component repulsion. The $R(\gamma )$ dependence is plotted for fixed $%
N=2.5$ and the potential's amplitude, $U_{0}=5$, hence the respective
effective mass, given by Eq. (\ref{meff}), is $m_{\mathrm{eff}}\approx $ $%
-0.34$. As shown in Ref. \cite{we1}, the $R(\gamma )$ dependence may be
accurately represented by means of a variational approximation. It is seen
in Fig. \ref{fig8}(b) that the antisymmetry of the two-component gap
solitons is broken at $\gamma <\gamma ^{\mathrm{(cr)}}\approx 0.28$. The
respective analytical result, determined by Eq.~(\ref{NC}) is $\left( \gamma
^{\prime }\right) ^{\mathrm{(cr)}}\simeq 0.272$, which is consistent with
the numerical findings.

\begin{figure}[h]
\begin{center}
\includegraphics[height=4.cm]{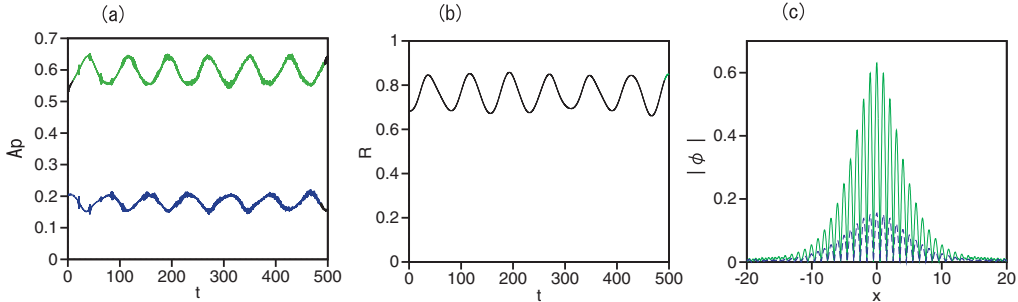}
\end{center}
\caption{(a) The evolution of maximum values of $|\protect\phi _{1}(x)|$ and
$\left\vert \protect\phi _{2}(x)\right\vert $ (green and blue lines), as
produced by direct simulations of Eqs. (\protect\ref{gap1}) and (\protect\ref%
{gap2}), initiated by the input in the form of ansatz (\protect\ref{phi12}),
with $\Phi _{1,2}$ taken as a numerically exact broken--antisymmetry soliton
solution of Eqs. (\protect\ref{gap3}) and (\protect\ref{gap4}) with $\protect%
\gamma =0.1$. (b) The evolution of the asymmetry measure for the solution
from panel (a), defined as per Eq. (\protect\ref{R}). (c) Snapshots of $|%
\protect\phi _{1}(x)|$ and $\left\vert \protect\phi _{2}(x)\right\vert $
(green and blue lines, respectively) of the solution from panels (b,c) at $%
t=500$. The snapshots closely approximate an exact shape of the
two-component gap soliton with broken antisymmetry between the components.}
\label{fig9}
\end{figure}

The prediction of the breaking of antisymmetry in two-component gap solitons
was confirmed by direct simulations of underlying equations~(\ref{gap1}) and
(\ref{gap2}), see an example shown in Figs. \ref{fig9}(a)-(c) for $\gamma
=0.1<\gamma ^{\mathrm{(cr)}}$. The initial conditions were taken as per
ansatz (\ref{Phi}):
\begin{equation}
\phi _{1,2}(x,t=0)=\Phi _{1,2}(x,t=0)\cos (\pi x),  \label{phi12}
\end{equation}%
where $\Phi _{1,2}(x,t)$ is the above-mentioned numerically exact
broken-antisymmetry soliton solution of Eqs. (\ref{gap3}) and (\ref{gap4}),
obtained by means of the imaginary-time simulations. Figures \ref{fig9}(a)
and (b) show, severally, the resulting evolution of maximum values of $|\phi
_{1,2}(x)|$, and of the asymmetry measure, $R$, defined according to Eq. (%
\ref{R}). Periodic small-amplitude variations of the fields and asymmetry
are caused by a deviation of ansatz (\ref{Phi}) from a numerically exact
form of the two-component gap soliton with broken inter-component
antisymmetry. A nearly exact shape of both components of the gap soliton is
produced in Fig. \ref{fig9}(c), which displays snapshots of $|\phi
_{1,2}(x)| $ at $t=500$. These numerical results also demonstrate that the
gap solitons with broken antisymmetry are stable. Similar to the structure
of the asymmetric localized states produced by Eqs. (\ref{1phi}) and (\ref%
{2phi}) in the presence of the HO trapping potential, see Fig. \ref{fig1}%
(a), the two-component gap solitons feature broken antisymmetry in the
central zone, and persistent approximate antisymmetry in decaying tails.

Thus, an essential difference from the results reported above for the
localized states, trapped in the 1D or 2D HO potential, is that the SSB
transition, revealed by Fig. \ref{fig8}(b), is categorized as an \textit{%
inverted bifurcation} \cite{bif} (alias an \textit{extreme subcritical
bifurcation}, cf. Ref. \cite{Dong}), in comparison with the supercritical
bifurcations observed in Figs. \ref{fig1}(c) and \ref{fig5}. The inversion
is explained by the fact that effective mass (\ref{meff}), which drives the
bifurcation of the gap solitons, is negative. Another obvious difference is
that the two-component antisymmetric gap solitons undergo SSB at $\gamma <1$%
, on the contrary to the above condition, $\gamma >1$, which is necessary to
impose SSB onto the HO-trapped states. On the other hand, it is known that,
in the interval of $1<\gamma <3$, the system of Eqs. (\ref{gap3}) and (\ref%
{gap4}) gives rise to a bifurcation which breaks symmetry of two-component
solitons with $\Phi _{1}=\Phi _{2}$ (the symmetric state, rather than the
antisymmetric one considered here) \cite{we1}. The consideration of the
latter effect in terms of Eqs. (\ref{gap1}) and (\ref{gap2}) is beyond the
scope of the present paper.

\section{Conclusion}

In numerous works, SSB (spontaneous symmetry breaking) of two-component
localized modes in linearly-coupled systems was found in the case of
attractive self- and/or cross-component nonlinear interactions in the
system. On the contrary to that, in systems with repulsive interactions
spontaneous breaking was only shown for antisymmetric two-component states,
but not for symmetric ones. Here, we have demonstrated that SSB in both 1D
and 2D symmetric states, trapped in the confining potential [taken as the HO
(harmonic oscillator)], is possible if the cross-repulsion is stronger than
intrinsic repulsion in each component. This setting may be realized in BEC
and nonlinear optics. In the former case, it represents a binary condensate
with natural repulsive contact interactions and radiofrequency-induced
linear mixing between two atomic states, which compose the binary BEC. In
terms of self-defocusing optical waveguides, the system is based on the
copropagation of two orthogonal polarizations of light with the linear
mixing induced by linear-polarization birefringence of the material.

For both one- and two-dimensional GSs (ground states), as well as for 2D
vortex states, the transition from symmetric states to asymmetric ones has
been demonstrated analytically by means of TFA (Thomas-Fermi approximation)
and confirmed by systematically collected numerical solutions of the
underlying system of linearly coupled GPEs (Gross-Pitaevskii equations). A
characteristic feature of the asymmetric states is that they combine
strongly broken asymmetry in the inner area, while a surrounding layer keeps
the original symmetry, in the approximate form. The SSB transition for all
these states is identified as a supercritical bifurcation. It produces
stable 1D and 2D asymmetric GSs, as well as stable asymmetric vortices with
topological charge $S=1$, while the vortices with $S=2$ are unstable against
splitting in a pair of unitary vortices.

The phenomenology of the spontaneous antisymmetry breaking was also briefly
considered for 1D antisymmetric two-component gap solitons, maintained by
the spatially periodic potential. In this case, the antisymmetry breaks in
the inner region of the gap soliton under the condition opposite to that
necessary for the occurrence of the SSB effect in the HO-trapped modes,
\textit{viz}., the cross-component repulsion must be weaker than the
self-repulsion in each component. The character of the antisymmetry-breaking
transition for the gap solitons is opposite too, namely, it amounts to an
inverted bifurcation.

The above analysis did not address motion of the trapped modes. Application
of a kick to the 1D trapped state, as well as of a radial push to 2D ones,
should excite a dipole mode of oscillations of the perturbed states around
the center. In this connection, the effect of the Rabi coupling on the
oscillations may be an interesting feature, as suggested by recently studied
effects of the same term on the motion of spinor solitons in a random
potential \cite{Mardonov}. Gap solitons feature mobility too, with a
negative dynamical mass \cite{we2}, and, accordingly, it may be interesting
to address an effect of the Rabi coupling on the motion of two-component gap
solitons.

This work can be also extended by considering higher-order (first of all,
dipole) spatial modes in the system with the HO trapping potential, as well
as antisymmetric and symmetric two-component configurations in the cases of
the HO and periodic potentials, respectively. A challenging possibility is
to develop the analysis for the three-dimensional version of the BEC model.

\section*{Acknowledgments}

The work of H.S. is supported by the Japan Society for Promotion of Science
through KAKENHI Grant No. 18K03462. The work of B.A.M. is supported, in
part, by the Israel Science Foundation through grant No. 1286/17, and by
grant No. 2015616 from the joint program of Binational Science Foundation
(US-Israel) and National Science Foundation (US).  This author appreciates
hospitality of the Interdisciplinary Graduate School of Engineering Sciences
at the Kyushu University (Fukuoka, Japan).

\end{document}